\documentclass[a4paper,man,natbib]{apa6}

\DeclareFixedFont{\ttb}{T1}{txtt}{bx}{n}{10} 
\DeclareFixedFont{\ttm}{T1}{txtt}{m}{n}{10}  

\usepackage{color}
\definecolor{deepblue}{rgb}{0,0,0.5}
\definecolor{deepred}{rgb}{0.6,0,0}
\definecolor{deepgreen}{rgb}{0,0.5,0}

\usepackage{listings}

\newcommand\pythonstyle{\lstset{
language=Python,
basicstyle=\ttm,
otherkeywords={self},             
keywordstyle=\ttb\color{deepblue},
emph={MyClass,__init__},          
emphstyle=\ttb\color{deepred},    
stringstyle=\color{deepgreen},
frame=tb,                         
showstringspaces=false            %
}}

\lstnewenvironment{python}[1][]
{
\pythonstyle
\lstset{#1}
}
{}

\newcommand\pythonexternal[2][]{{
\pythonstyle
\lstinputlisting[#1]{#2}}}

\newcommand\pythoninline[1]{{\pythonstyle\lstinline!#1!}}

\usepackage[english]{babel}
\usepackage[utf8x]{inputenc}
\usepackage{amsmath}
\usepackage{graphicx}
\usepackage[colorinlistoftodos]{todonotes}
\usepackage{paralist}

\title{Social Bots: Human-Like by Means of Human Control?}
\shorttitle{Social Bots: Human-Like by Means of Human Control?}
\author{Christian Grimme, Mike Preuss, Lena Adam, and Heike Trautmann\\ \{christian.grimme, mike.preuss, lena.adam, trautmann\}@uni-muenster.de}

\affiliation{University of Münster\\Department of Information Systems\\Leonardo-Campus 3\\48419 Münster, Germany}

\abstract{Social bots are currently regarded an influential but also somewhat mysterious factor in public discourse and opinion making. They are considered to be capable of massively distributing propaganda in social and online media and their application is even suspected to be partly responsible for recent election results. Astonishingly, the term `Social Bot' is not well defined and different scientific disciplines use divergent definitions. This work starts with a balanced definition attempt, before providing an overview of how social bots actually work (taking the example of Twitter) and what their current technical limitations are. Despite recent research progress in Deep Learning and Big Data, there are many activities bots cannot handle well. We then discuss how bot capabilities can be extended and controlled by integrating humans into the process and reason that this is currently the most promising way to go in order to realize effective interactions with other humans.

}

\begin{document}
\maketitle

\tableofcontents


\section{Introduction}

Social media is a phenomenon that exists for a bit more than a decade now (Facebook went online 2004, Twitter in 2006). For the first time, a large part of the world population is enabled to participate in direct and partly world-wide visible information exchange.

Together with the increasing importance of the social media in all-day live and a growing reach of these networks, their use (or misuse) for orchestrated information distribution in terms of advertisement up to political propaganda becomes attractive for different stakeholders. Due to the underlying technical nature of the communication medium, automated and thus cost efficient access to social media channels is easy. Like for email services several years ago, social media channels are used for simple spamming~\citep{Tynan2012}. However, since about 2010, reports on trolling or automated so-called social bot activity in social media increase - especially with a focus on political manipulation and propaganda~\citep{Chu2010,Fredheim2013,Elliott2014}. Today, it is not doubted that social bots have a high societal impact \citep{NatoSocialBotsReport2017}, whatever the approaches realizing them currently are.
This leaves research with new and multidisciplinary challenges: Detecting and fighting automated and orchestrated manipulation via social media necessitates insights and understanding of motivation, processes, economics, and current limits of manipulation. Computer scientists track networks, measure interactions, build algorithms, and are concerned with security issues, but are unfamiliar with communication aspects and effects. Social scientists have to understand new (semi-automatic) ways of distributing information or propaganda and answer questions of possible societal impact. Both have to collaborate with statisticians and researchers in the area of artificial intelligence to understand challenges and limits of developing big data-based detection mechanisms.

As a first step, this work covers technical details and processes, economic considerations, as well as limits of automated manipulation via social networks in a multi-disciplinary way and provides some baseline for further discussion. For an initial common understanding, we review the existing interpretations of the term "social bots" and propose a consolidated definition. This definition is complemented by a comprehensive discussion and classification of automated actors in the web. 

Then, we focus on the technical details and challenges in the development of social bots. We first show the construction and implementation principle of a responsive Twitter bot and extend this implementation to a framework for realizing human-like behavior. Additionally, the second concept is validated by a social bot experiment at Twitter, applying 30 of our social bots for gaining followers and distributing (ethically harmless) content. For both concepts we discuss the costs of realization. 

In a final step, we address the existing gap of automation on the behavioral level and automation on the communication level. Here, we argue that it is currently most cost efficient to automate bots on a behavioral level, while content generation and bot-human-communication is still steered by humans. In the context of the performed bot network experiment, we empirically show that current automatic detection mechanisms cannot significantly distinguish hybrid bots from human users.

\section{Definition and Taxonomy of Social Bots}
\label{sec:definition}

When journalists, bloggers, or scientists report on social bots and their potential influence on society, many of these articles provide an own definition of the term "social bot". Very often, these definitions strongly differ from each other, some focusing on technical details, others highlighting social interaction. Sometimes, the definitions even contradict each other or explicitly exclude a class of "social bots" others include. Although the capabilities and effects of social bots taking part in public internet communication are more and more discussed, no common understanding of the vehicle itself has evolved.

As the term itself suggests, definitions stem from a mixed, partly social science and partly technical perspective, while the weighting of the perspectives is usually up to the respective definition's author.

From a technical perspective, the term "bot" is often related to robots, automation and algorithms~\citep{Marechal2016}. All of these terms are certainly part of the understanding of social bots, however, their equivalent substitution interweaves technically different concepts like algorithms and robots in a simplistic way and may lead to misunderstandings. \citet{Geiger2016} defines social bots---in a more exact but still very general way---as automated software agents. Emmer (interviewed in \citet{Boell2017}) adds properties like artificial intelligence and the ability to autonomously act in the web.

The social science perspective usually addresses the social or political implications of the actions of social bots. \citet{Woolley2016} states that social bots "mimic human social media users" and "manipulate public opinion and disrupt organizational communication". He also defines so called "political bots" as a special case of social bots. \citet{Hegelich2016} highlights that social bots are hidden actors with a political agenda. He explicitly distinguishes them from "chat bots" or other "assistants". A wider definition which specifically considers the communication behavior is given by \citet{Frischlich+2017}. The authors point out, that the imitation of human communication (behaviour) is a key feature of social bots. This certainly also includes chat bots. Even more general, \citet{Kollanyi+2016} consider interaction with other users through automated social media as key property of social bots. Interestingly, social media platforms like Facebook recently recognized possible effects of social bots by admitting "false amplifications", however, they do not use the term "social bot" throughout their publication~\citep{Facebook2017}. 

Many application examples of social bots are presented in a recent overview paper by \citet{Ferrara2016}. This work allows to identify many types of bots and check the available definitions. Additionally, the authors give a good but (in our view) slightly too tight definition of social bots: "A social bot is a computer algorithm that automatically produces content and interacts with humans on social media, trying to emulate and possibly alter their behavior." We will keep several aspects of this definition but do not restrict ourself to social media alone. Additionally, we include the communication aspect introduced by \citet{Frischlich+2017} and cover the interaction property by referring to agent behavior:\\[0.2cm]

\textbf{The term "Social Bot" is a superordinate concept which summarizes different types of (semi-) automatic agents. These agents are designed to fulfill a specific purpose by means of one- or many-sided communication in online media.}\\[0.2cm]

The most significant difference to other definitions is that we define social bots as a high-level concept which comprises many types of specific bots. Additionally, our definition covers:
\begin{compactitem}
	\item fully automated as well as partly human-steered bot action,
	\item autonomous action (agent-like),
	\item an orientation towards a goal,
	\item multiple modes of communication,
	\item and a wider ecosystem (all online media).
\end{compactitem}

In the following paragraphs, we give several examples of social bots and specific sub-types as well as of bots which are not covered by our definition and, thus, are not supposed to be social bots.

\subsection{Social Bots}
The most popular type of a social bot is the chat bot, „a software system, which can interact or “chat” with a human user in natural language such as English.“~\citep{shawar2007chatbots}.  The oldest and best known chat bot may be Joseph Weizenbaums ELIZA. It was able to participate in a discussion on psychological topics, controlled by scripts that discover context by identifying keywords~\citep{weizenbaum1966eliza}. By means of pattern matching, ELIZA answered questions in a very human like manner, so sometimes participants did not even recognize that they talked to a machine instead of a real therapist.
The recent wave of chat bot development probably originated  in the context of the “Loebner Prize” competition\footnote{\url{http://www.loebner.net/Prizef/loebner-prize.html}}, where Hugh Loebner set the task to find the most human-like acting program~\citep{mauldin1994chatterbots}.

Nevertheless, chat bots are only as intelligent as their scripts and the databases behind those scripts are. That is why they are often developed only for specific topics.
Nowadays, lots of different chat bots with different aims exist \footnote{\url{https://www.chatbots.org}}. Meanwhile, companies often use chat bots to handle customer service issues.  One can find them “in daily life, such as help desk tools, automatic telephone answering systems, tools to aid in education, business and e-commerce.“~\citep{shawar2007different}. As chat bots are created to communicate in dialogs with specific users or customers, a multitude of chat platforms are conceivable, e.g. private chats of social media pages, as well as other online media like email or help sections on private company websites. The bots partly replace human interaction and are often used to do simple preprocessing tasks, e.g., figuring out the right contact person for a specific service issue. 

Whereas chat bots focus on one-to-one communication, spam bots are developed to reach a large audience. The goal of this one-to-many communication is spreading information, advertisements or fishing links, without involving the recipient. As they were used to communicate a certain message on behalf of a company, group, or person, they nevertheless fall into the category of social bots.

As mentioned earlier, political bots can be seen as a special type of social bots with the aim to spread political content or participate in political discussions on online platforms~\citep{Woolley2016}. Political bots are designed by politically oriented groups to represent their opinions and mindsets. A typical goal of political bots is, e.g., boosting the popularity of a specific idea or person on a (social) media platform, by generating ‘likes’ or ‘follows’.  Furthermore, political bots may make use of the characteristics of chat bots or spam bots. They discover public conversations, posts and comments by identifying keywords and intervene or flood them with own (propagandistic) content. Whether political bots are able to participate in simple conversations with other users or just spread spam in a not reactive way is just a question of the aim (and technical skill) of the operator and the code behind the bot profile. Human-like political bots that act on social media platforms as Twitter and Facebook or comment in forums are potentially capable to influence other users. Especially, if there are many bots cooperating in bot networks, they are able to arise a potentially undeserved awareness to political moods. Examples of potential bot armies were, e.g., discovered in the context of the U.S. presidential election 2016. ~\citet{bessi2016social} found out that nearly 19 percent  of all election-related Twitter posts during this time were made by bots. Furthermore, the German news page "Spiegel Online" reports on parties which considered the usage of social bots supporting their election campains for the German parliamentary elections of 2017~\citep{rosenbach_online_2016,pfaffenzeller_online_2017}. 

Another type of social bots is the class of mobile phone assistants. Software like Apples Siri\footnote{\url{https://www.apple.com/de/ios/siri/}} is designed to manage human to machine communication with the input of natural language. Nearly any possible functionality of the mobile phone can be used just by voice commands. In this case, the social bot acts as a translator between human users and the phone. With the help of voice recognition and keyword identification, the program figures out appropriate actions or search results for the user. 


\subsection{Bots Not Regarded as Social Bots}

Bots that are not covered by our definition of social bots are, e.g., content management bots, aka ‘curator bots’. The job of a curator bot is to manage or collect content and to present it in an easy-to-digest way to humans. In contrast to social bots, for curator bots the communication aspect is not pronounced; they only work ‘silently’ with content. Wikipedia bots are an appropriate example for this class of bots. Pywikibot\footnote{\url{https://www.mediawiki.org/wiki/Manual:Pywikibot/Overview}} helps users to nurture articles by deleting superfluous whitespace, generating links to related pages or correcting typos. Another example of content bots are data aggregation bots which are built to manage data and are used for analysis only. 

Game bots help their users to be successful in games. Tasks of these bots can be as various as the games they are used in. Game bots can act as opponents in order to enable training, help to navigate through the game or can be used for cheating or just as stand-in for short periods of unavailability (afk). So-called farming bots as e.g. in games like World of Warcraft\footnote{\url{https://www.worldofwarcraft.com/}} assume simple tasks and free players from time-consuming but necessary duties~\citep{mitterhofer2009server}. 
Nowadays, game bots that realize all these functions and more are available in USB stick format from graphics card vendors.
Instead of social bots, game bots focus not on communication and interaction but exclusively on substituting users by imitation.

Service-Level-Agreement (SLA) negotiators focus on machine-to-machine communication. These bots are built to handle Service Level Agreements autonomously. 
Again, there is no human communication or interaction aspect regarding this class of bots, which is why they are not covered by our definition of social bots.


\subsection{Discussion}
As also shown by the categorization into social and not social bots above, we see the human-machine interaction as a key-factor. Social bots automate social interaction via communication
Every online medium where human communication through publicly visible posts, chats, comment-functions, direct messages, etc., takes place, is a possible point of connection for the involvement of social bots. Nevertheless, our definition of social bots should be seen more or less as a high-level concept. Social bots appear as different from each other as the reasons they were built for, and have to be discussed in their specific context. Having a look at the mentioned examples, one can see that there are more and different tasks for social bots than to influence people. Some social bots just substitute their users by assuming duties or handle simple preprocessing tasks. Announcements as Facebook's support for group bots and bot repositories\footnote{\url{https://techcrunch.com/2017/03/29/facebook-group-bots/}} lead us to expect that social bots are going to be a more and more pervasive part of our internet experience in the coming years. 
When discussing the possibilities of social bots to influence single users up to whole societies, we shall therefore employ more precise notions and terms.




\section{Automation using Social Bots}
The application of social bots for multiple purposes (from advertisement to propaganda) implies different technical challenges as well as economic considerations to be handled. On the one hand, costs are rapidly increasing in terms of technical complexity (e.g. for making social bots more  human-like). On the other hand, simple technical realizations may have a big enough impact to maximize monetary or social/political revenue in some cases. In the following sections, we will present a most simple technical realization of a social bot and extend it to a moderately complex behavioral human-like actor on Twitter---trying to keep costs rather low. We then present an experiment using 30 of those Twitter bots and lead over to an economic discussion of hybrid extensions of social bots in the next chapter.

\subsection{A simple reactive Twitter Bot example}
\begin{figure}[h!tb]
	\centering
	\includegraphics[width=0.7\textwidth]{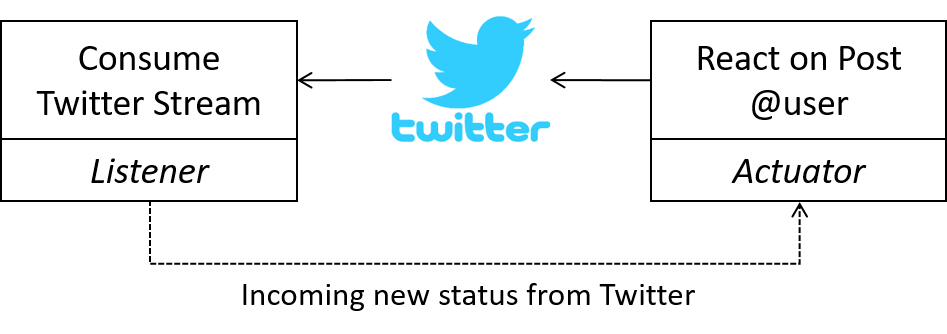}
    \caption{\label{fig:simple_bot}Components and data flow of simple Twitter bot realization using the Twitter Stream API.}
\end{figure}

One of the most simple ways to develop a reactive social bot adopts the Twitter Stream API\footnote{\url{https://dev.twitter.com/streaming/overview}}. This basically means, that we listen to the ongoing worldwide Twitter activity and react to arriving posts. More formally, we use a Twitter Stream Listener component that registers with Twitter and additionally implement a simple actuator component which is triggered by incoming Twitter posts and uses the Twitter REST API\footnote{\url{https://dev.twitter.com/rest/public}} to reply to these posts, if applicable. Figure~\ref{fig:simple_bot} provides a schematic overview of the components and data flow of the social bot. For the full implementation details using the Python \texttt{tweepy}\footnote{\url{http://www.tweepy.org/}} framework, refer to the listing given in appendix~\ref{sec:simple_bot_listing}. Note that due to bandwidth management of the public Twitter Stream, only a subset of posts will reach the listener. Depending on the registered topics and activity at the platform, only about 1\% up to 40\% (in very restricted cases also more) of the actual traffic may arrive at the listener~\citep{Ferrara2016}.

\subsubsection{Functionality}
The presented social bot enables us to react on twitter posts directly, answering to the sender. In our implementation, the Twitter Stream listener consumes the current Twitter Stream with respect to a given set of hash tags or topics. Thereby, we are able to adapt to a specific context or domain of interest. Although the current implementation only greets the user of current post, the functionality of the actuator can easily be extended. The application of this bot ranges from simple demonstration (and greeting) purposes to simple service activities based on standardized responses, e.g.:

\begin{compactitem}
	\item Returning the weather forecast for a city or region mentioned in the current post. Therefore, the actuator may use external weather information sources like OpenWeatherMap\footnote{\url{https://openweathermap.org/api}}.
    \item Answering questions on specific topics detected in the current post. Using the Google Knowledge Graph\footnote{\url{https://developers.google.com/knowledge-graph/}}, a mighty ontology network can be connected to the bot, covering an enormous knowledge base.
    \item In a political context: Respond to specific topics and confront users (usually independent of content posted) with a number of fixed political statements.
\end{compactitem}

Only these three application examples demonstrate the potential of a very simple social bot implementation comprising not more than 30 lines of code for a fully functional frame.

\subsubsection{Costs}
Obviously, the costs for developing a simple service social bot can almost be neglected. Implementation time is certainly lower than one hour for a medium experienced developer (including error handling code, which is not provided in our listing). The main effort has to be put into the setup for the bot's Twitter account and access to the Twitter API. Therefore, a standard Twitter account must be created and connected to a mobile number for developer access. Both can easily be done in an anonymous way using a fake email address and an invalid or anonymously registered mobile number.

Clearly, the behavior of the presented bot can easily be detected as automated action. No human recipient of a message will consider it to be sent by another human. Instant reaction, permanent activity, and restricted capabilities to analyze content and react dynamically will expose the social bot as such.

\subsection{A Social Bot with human-like behavior}
Development of a social bot with sophisticated human-like behavior addresses three main challenges:
\begin{compactenum}
	\item Producing credible and "intelligent" content, which is accepted as such by human consumers.
    \item Leaving a trace of human-like meta-data in social networks.
    \item Creating an adequate (often balanced) network of friends or followers to spread information.
\end{compactenum}

While the first challenge is a rather open issue in science and even the more in practice (we will comment on this later), the second aspect can be handled to a certain extent by imitating human actions in social networks sticking to normal human temporal and behavioral patterns. This includes performing activities in a typical day-night-cycle, carefully measured actions at the social media platform, as well as variability in actions and timing. Thus, at Twitter, a bot should pause actions to simulate phases of inactivity (e.g. sleep or work), limit posting and re-tweeting activities to a realistic, human-like level, and vary these pauses and limits. 

Another key issue is to grow a network of followers or friends. For social media, a network of friends implies a certain reach: the larger the network of followers, the more Twitter users receive distributed content of the respective account. 
To create a network, \citet{Lehmann2013} proposes an effective strategy based on a simple observation: users follow other users hoping that those follow back again (which they often do, if the pro-active profile does not obviously look bot-like), thereby  establishing a friendship relation. In case this does not happen within a certain time span (i.e. the other user does not follow back), the one-sided connection is often dissolved to keep a balanced following-follower-ratio. An exception to this are very prominent accounts with usually strongly imbalanced following-follower-ratios (far more followers than followed users) or accounts that are mainly used to distribute advertisement (far more followed users than followers). 
These clearly indicate not human-like behavior and are used for bot-detection. The overall principle of "follow-for-follow" is not only respected by most human users but can also be applied to grow the follower network of a bot account.

\subsubsection{Extending bot functionality}
\begin{figure}[h!tb]
	\centering
	\includegraphics[width=0.9\textwidth]{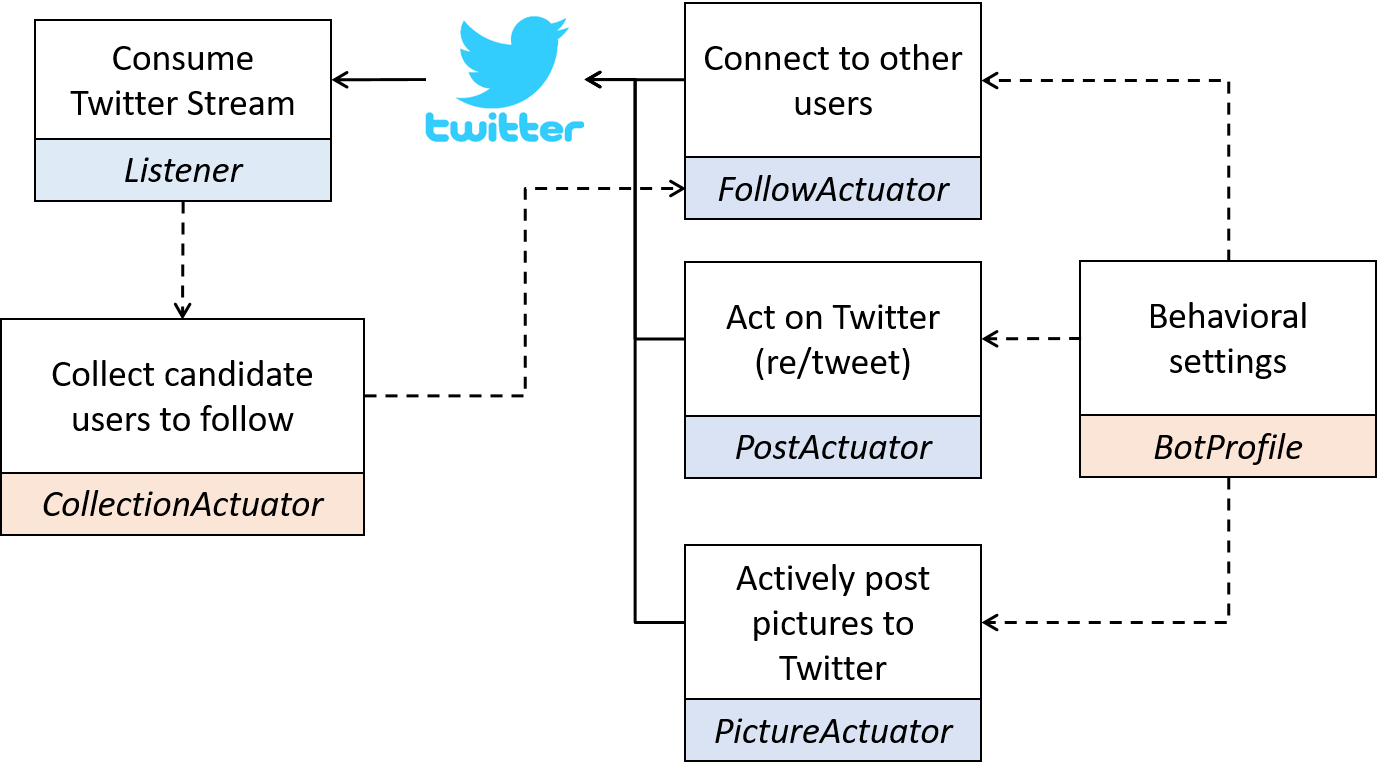}
    \caption{\label{fig:exp_bot}Components and data flow of our advanced social bot with behavioral settings, follow-for-follow mechanism, and human-like activity profile.}
\end{figure}

The previously presented simple social bot can easily be extended to fulfill the challenges two and three. Therefore, several Actuators are created that independently perform specific actions on Twitter. Considering the schematic depiction in Figure~\ref{fig:exp_bot}, we briefly describe the important components:
\begin{description}
	\item[CollectionActuator:] This component listens to the Twitter stream and stores user names as candidates to follow later on. The selection of following candidates can be made with respect to different characteristics like the following-follower-ratio (i.e. balanced accounts are preferred), activity on Twitter (potential multiplicators are preferred), Tweet properties (e.g. users sending popular tweets are preferred).
    \item[BotProfile:] The personal profile of a social bot is defined using a dedicated component which stores all constraints and guidelines to simulate a certain behavior. Here, the day-night-cycle and rest periods can be defined, general parameters for the posting and re-tweeting behavior can be set, and an individual following behavior is formulated. Note that all settings should be guidelines only, in order to add some random variability. The component also provides functionality to request the next action time for all other actuators. This function interprets the given behavior values and (adding some random noise), proposes the next action.
    \item[FollowActuator:] This actuator ensures a continuous execution and management of the follow-for-follow procedure. With respect to the BotProfile, the component follows a certain amount of previously collected users (see CollectionActuator) and supervises reactions. Is a contacted user follows back, the component adds this user as friend. In case of no response within a certain time window (i.e. 24 hours), the one-sided friendship is canceled and the user is blacklisted.
    \item[PostActuator:] This Actuator enables the bot to post or re-tweet on Twitter. Therefore, a database of individual tweets and collected tweets is accessed. The amount of actions is determined by the BotProfile.
    \item[PictureActuator:] The ability to post pictures on Twitter is implemented by this Actuator. Analog to the behavior of the PostActuator, pictures and matching comments are extracted from a picture database and posted on Twitter.
\end{description}

\subsubsection{Experimental evaluation}
In order to evaluate our mimicry approach for human behavior in the real-world context, we set up an experiment comprising 30 Twitter bots. In cooperation with the German TV station Pro7, we created 30 fake profiles, see also Table~\ref{tbl:bot_detection}, and equipped them with the social bot framework described before. Each bot ran the same code, however, we individualized the bot profiles and Twitter Stream listeners. Each social bot had its own day-night-cycle, activity pattern, and following behavior. Additionally, each bot listened for an individual set of topics within the Twitter Stream. Overall, the experiment was divided into three phases:
\begin{enumerate}
	\item Build a network of followers over a setup and testing period of 2 days and the following 8 days of combined action. Thus, the experiment lasted 10 days in total of which only the 8 productive days were documented. Note that half of the social bots were mutually befriended by default, while the second half started with no followers at all.
    \item Publish content in a coordinated way to test the potential of setting a trend on Twitter. The published content was devised by human actors and only distributed by the bots.
    \item Reveal the social bot identity of the respective fake accounts to followers and the public (supported by a TV documentary on the experiment, which is available in German\footnote{\url{http://www.prosieben.de/tv/galileo/videos/2016347-social-bots-das-experiment-clip}}).
\end{enumerate}

\begin{figure}[htb]
	\centering
    \includegraphics[width=.7\textwidth]{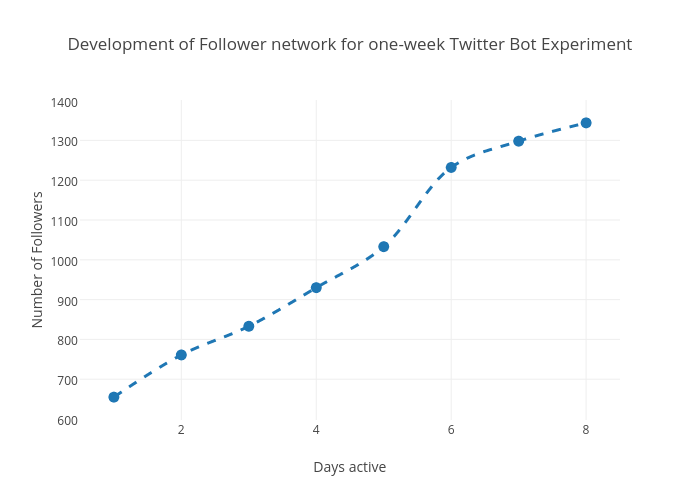}
    \caption{\label{fig:bot_followers}The plot shows the growth of the follower network for the initial Twitter bot setup in about one week. 27 of initially 30 Twitter bots continuously performed the follow-for-follow strategy automatically without any human intervention. Potential followers were selected from the Twitter stream regarding individual topics.}
\end{figure}

Especially phase one demonstrated, that the follow-for-follow approach could successfully be applied to acquire followers automatically. As shown in Figure~\ref{fig:bot_followers}, the amount of followers continuously increased for the evaluated eight days, resulting in about 1350 followers after this short time period.
During the second phase, two (harmless and humorous) hash tags were promoted to test the reach of the acquired followers. Although the hash tag briefly appeared in the German top 100 trending topics, a significant trend could not be established.
However, phase three showed, that many human followers had been deceived by the fake bot identities and actions. Reactions from Twitter users were different ranging from disappointment to anger and from amusement to disbelieve.

Although our experiment is only a snap-shot of what is possible by applying human-like acting social bots, some important insights can be extracted:
\begin{itemize}
	\item Tedious tasks as building a follower network, as well as posting and re-tweeting content, may be automated without being exposed as bot.
    \item The automatically generated network can be used to spread content to all followers at any point in time. This will cause at least brief visibility and possibly push a topic in order to reach wider popularity.
    \item Human users can easily be deceived by simple, but fairly realistic social bots behaviors.
\end{itemize}

Certainly, an important ingredient for the success of our social bots was---besides the human like behavior patterns---the human-generated content published by all bots. As mentioned before, we used manually generated content to be spread by the bots. We include the discussion of this aspect into our cost review.

\subsubsection{Costs}
The development time of the extended Twitter bot (less than two days) can still be neglected compared to the functionality and benefit of automation provided by the general framework. The more tedious task was to generate all thirty fake accounts on Twitter. Thereafter, we were able to deploy the same code thirty times with only minor adaptations regarding the individual configuration of each bot. Then, phase one (growing the network) was performed by all bots without any human intervention.

Likewise, publishing content in phases two and three needed no intervention. However, content was not automatically generated but provided by humans. We decided to do so after reasoning on the following to questions:
What would have been the costs of generating content automatically, and what quality of content can be achieved?

Implementing the generation of {\it intelligent} and {\it creative} content for our hash tags would have cost far more effort than setting up the whole social bot framework. Simple approaches based on templates still require some human interaction and lack creativity. More complex generators based on learned patterns still follow firm rule sets, which limits the variability of linguistic expression. Both probably would have had reduced the credibility of our social bots due to repetitive content. Furthermore, due to the application of thirty cross-linked social bots and their continuous re-tweeting behavior, a single message was repeated many times by other bots and followers, thereby extending its range automatically.

\section{Hybrid Social Bots}
\begin{figure}[h!tb]
	\centering
	\includegraphics[width=0.9\textwidth]{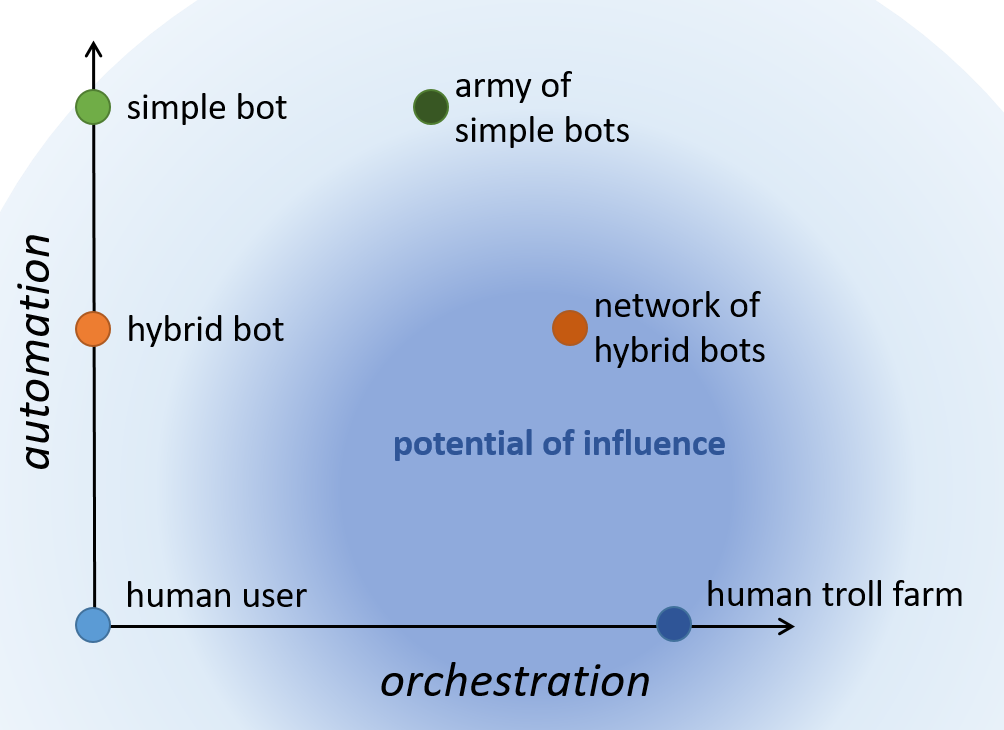}
    \caption{\label{fig:auto_orchestration}Qualitative classification of the potential influence of humans and bots in social media with respect to automation and orchestration.}
\end{figure}

The extended social bot framework presented in the previous chapter is able to mimic human behavior on the action level, i.e. a social bot is able to automatically create a follower network, and manage content. Content production, however, is done by human actors. Figure~\ref{fig:auto_orchestration} qualitatively shows the automation-orchestration relationship of human users and simple social bots as single actors and as human troll farm or "bot army" respectively. Hybrid social bots are, with respect to automation, an intermediate class of fully automated (behavioral simple) bots and purely human users. Used under orchestration, communication approaches and activity patterns of single actors differ: The army of simple bots is often following a mere client server model with rather similar acting single bots and little autonomy per agent. Hybrid bot networks are certainly still centrally controlled. Each bot, however, possesses a behavioral autonomy, which mimics human behavior. In contrast, Human troll farms are acting on a central interest, context or overall goal but have the highest autonomy per agent. For them, a central content generation becomes dispensable.

In this chapter, we argue that hybridization of bots is an effective (compared to an army of social bots) and low-cost (compared to a human troll farm) approach to gain a high potential of influence via social media by simulating human behavior and speech. We will show that a network of these hybrid bots is able to sufficiently outsmart current automatic detection mechanisms as BotOrNot\footnote{\url{https://botometer.iuni.iu.edu}}~\citep{varol+2017}.

\subsection{Hybridization as low-cost mimicry approach}


The current societal opinion on bot technology seems to be driven by recent success stories of AI, for example the prominently featured wins of an AI against world-class Go players. These successes follow to a large part 
from the development of deep learning algorithms \citep{LeCun2015} that are a) able to employ big data collections for learning and b) benefit from extreme parallelization. However, at the core of deep learning successes, we see human competitive (or even better) pattern matching and modeling capabilities. It is not at all trivial to use these capabilities to establish creative tasks, and especially human communication skills are still beyond of what algorithms can do.

\cite{Riedl16} gives an overview and vision of how AI can approach computational interactive narrative, which requires that computers can understand human communication and react adequately. Attempts in this direction are currently still very limited, as, e.g. shown by \cite{Martin2016} in terms of computational improvisation in relatively open (not targeted) communication. 

Existing chat bots can answer simple questions in a limited domain of their expertise but lack skills to participate in an open discussion. Recent attempts to improve their capabilities include the 
ParlAI\footnote{https://code.facebook.com/posts/266433647155520/parlai-a-new-software-platform-for-dialog-research/} platform published by Facebook, but these approaches are currently active research directions.
Whereas progress in modification of images is astonishing (\cite{ZhuPIE17} provides a tool that can translate images to other styles, e.g., the style of a specific painter), this is not yet possible for working with text, which is, in this respect, considered much more complex than image translations.

In the related fields of computational creativity and procedural content generation (mainly for games), we see similar problems, which has led to so-called mixed initiative approaches \citep{Liapis2016} where a human designer and a computer program work together, taking turns, in order to reach a specific design goal. Without human interaction, the available methods would not be able to produce results in a human compatible style.  
At least for the time being, it is seemingly mandatory to employ hybrid approaches in order to establish results that can be taken for generated by a human and thereby appear human-like.

\subsection{Hybridization as strategy against rule-based detection mechanisms}
In order to evaluate our social bots against state-of-the-art detection mechanisms, we confronted them with the BotOrNot service provided by Indiana University~\citep{varol+2017}. The BotOrNot service tries to state on the overall probability that a submitted Twitter account is automated. Therefore, the service compares previously learned patterns regarding the account's meta data, network, behavioral timing, friendship relations, sentiment, and content. The authors report of more than $1,150$ features that constitute the patterns in all the named high-level classes. Finally, the results of all indicators are aggregated to a value in $[0,1]$ which represents a probability of an account being controlled by a social bot.
Table~\ref{tbl:bot_detection} shows the overall rating for each continuously active social bot account of our experiment. Obviously, the probability ranges between $0.37$ and $0.6$ with an average of $0.48$. That confirms, in average, no clear bot-identification is possible for our social bots.

In order to judge on the quality of these score distribution for our bots, we generate a baseline distribution of score BotOrNot values of worldwide user accounts. 

\subsubsection{Methodology}
As basis for user extraction we used data from the Twitter Decahose Stream, which provides a random 10\% sample of worldwide Twitter traffic. The Twitter Decahose Stream provides roughly 300 posts  per second. This sums up to about 160 GB of data per day. From this huge data sample of a single day, we extracted unique user accounts at four points in time: midnight, morning (6 am), noon (12 am), and evening (6pm) to respect possible effects of the day-night-cycle. The gathered user accounts (about 1200) were classified by using the BotOrNot-API provided by the BotOrNot service. As our Social bots acted in the German language domain, we additionally extracted only German user accounts at the same points in time for a second, localized baseline distribution of scores.

\subsubsection{Comparison of Bots and average accounts}
\begin{figure}[htb]
	\centering
    \includegraphics[width=\textwidth]{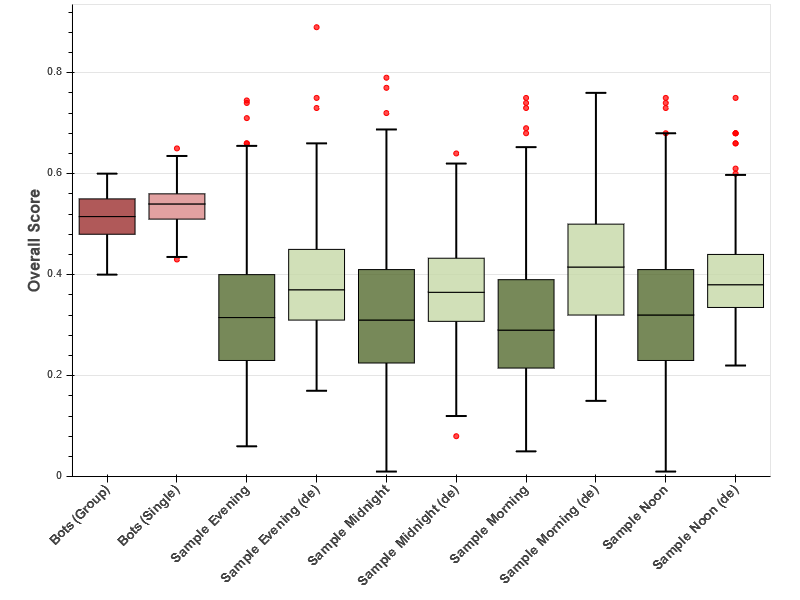}
    \caption{\label{fig:bot_overall_score}Statistics of the overall BotOrNot score values for our the social bot network (red box plots, grouped into bots that initially act as single entity or group respectively) contrasted with two baseline overall scores for a set of sample users. The sample users are taken from the worldwide (green box plots) and German (light green box plots) Decahose Twitter stream at four points in time. The analysis was performed using the BotOrNot API.}
\end{figure}

The comparison of our social bots' overall scores to the baseline distributions for the worldwide and and German users is shown in Figure~\ref{fig:bot_overall_score}. Although the bots cannot clearly be classified as bots with respect to the score measure, in retrospective evaluation, their score is significantly higher than the baseline score of our sample score from the worldwide and German Twitter Stream.

To further analyze these findings, we additionally take a look at the detailes meta-features provided by BotOrNot and the according scores.

\begin{figure}[htb]
	\centering
    \includegraphics[width=.4\textwidth]{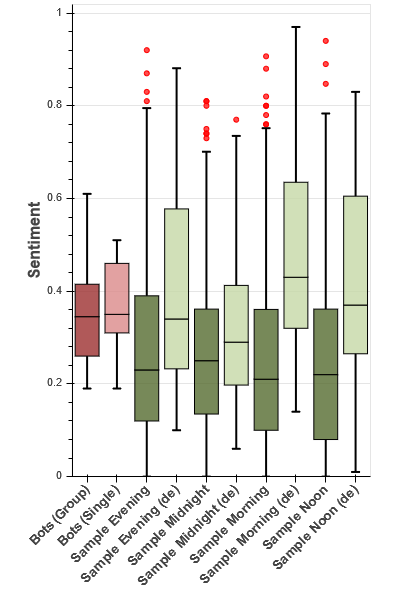}
    \includegraphics[width=.4\textwidth]{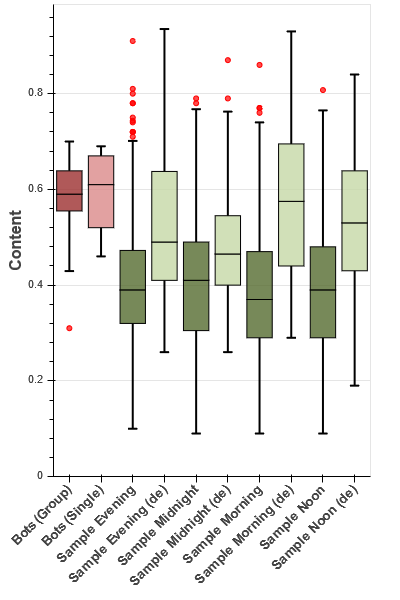}
    \includegraphics[width=.4\textwidth]{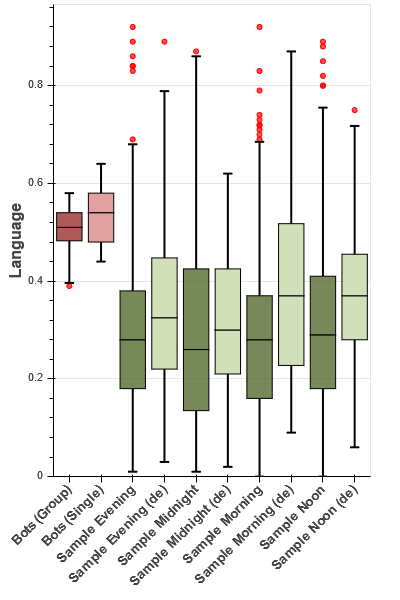}    
    \caption{\label{fig:bot_score_detail1}Detailed statistics of three meta features (sentiment, content, and language) for our social bots (red) and the baseline accounts worldwide (green) and from Germany (light green).}
\end{figure}

\begin{description}
	\item[Content-related features:] Figure~\ref{fig:bot_score_detail1} shows the detailed results for the sentiment, content, and language scores. For the sentiment score features like happiness, valence, arousal, and dominance as well as polarization and emoticon statistics of tweets are evaluated and aggregated. Here our bots obviously behave like baseline German users. Both, German users and bots are generally scored higher than the worldwide baseline, which may be caused by the fact, sentiment analysis for the German language is more difficult than for e.g. English. The same observation can be made for the content feature, which aggregates tweet length and entropy. Here, the bot also range in the German baseline. Language features combine statistics on part-of-speech tags in tweets, i.e. low level features on the tagged or annotated grammar and context of words used in the tweets. Here, a significant difference to both baselines is observed. The reason for this may be the high amount of slang terms and thus grammatically complex structure of tweets used to push a topic in phase 2 of our social bot experiment.
    \item[Meta data-related features:] For meta data features we observe that our bots behave in average similar to the German user baseline, except for the user score. The user score aggregates account-specific meta data information like age of the account and profile description as well as frequency and temporal development of actions on Twitter. Especially for these features, our experimental bot accounts are certainly too short-lived to be classified human-like. All other meta-features, however, confirm human-likeness of the social bot behavior---especially, when we consider friendship features, networking and temporal behavior. Here, no significant difference to the German baseline accounts can be identified.
\end{description}

\subsubsection{A comment on detection mechanisms and hybridization}
The evaluation of our social bot network showed that multiple features of a state-of-the-art detection tool like BotOrNot can be bypassed. Especially the scores "attacked" by our automation framework (friendship, network, temporal behavior) are not distinguishable for bots and the evaluated random German account sample. Only features on content and the user profile showed some indications for bot behavior. These indicators, however, are only identifiable due to an a-priori grouping of the known social bot accounts. If confronted with a single bot account, the BotOrNot detection mechanism does not provide a sufficient overall score to identify any of our social bots as such.

\section{Conclusion}
With this paper, we have contributed an interdisciplinary perspective on social bot taxonomy, degrees of automation, developmental costs, and the benefit and importance of human interaction for making social bots invisible for modern detection mechanisms. In detail, we gave a consolidated definition of social bots and applied it to known variants of automated actors in the web. From a more technical perspective, we provided insight into the implementation and costs necessary to deploy simple but reactive social bots in Twitter. To increase credibility, we extended the simple bot implementation by mimicking human behavior in temporal and operational properties. Content production was left to human controllers leading to a hybrid bot network. We experimentally deployed such a network and demonstrated its principle applicability. Tedious tasks were automated (like collecting followers, re-tweeting, or posting human-prepared content). Finally, we discussed the costs and current technological limits for full human-like hybridization. Futhermore, by means of  an empirical analysis from the Twitter bot experiment and average user data extracted from the Twitter Decahose Stream, we have shown that hybrid social bots are able to bypass important indicators of current rule-based detection mechanisms as BotOrNot.

Our results reveal several new challenges for future research in social bot detection: The next big challenge for detection systems will be to identify hybrid social bots, which expose real human behavior, on the one hand, and automatic patterns in some actions, on the other hand. We assume, that rule-based methods will not suffice for these tasks. In fact, adaptive and real-time detection mechanisms, which are able to reconfigure and learn are necessary to react on changing behavioral patterns almost instantly. Additionally, we believe that the inclusion of human interaction into hybrid social bots should shift the focus from purely automatic detection systems to hybrid detection systems that are able to judge on content, background strategies and distributed narratives by the inclusion of human (possibly crowd) intelligence.

\begin{table}[htb]
\caption{\label{tbl:bot_detection} List of all Twitter bots used during the experiment including the probability of being a Twitter bot determined by BotOrNot for each account.}
\footnotesize
\begin{tabular}{llr||llr}
\hline
Bot name & Bot Twitter ID & BotOrNot & Bot name & Bot Twitter ID & BotOrNot\\\hline
MagaritaWolff & 803538518014300160 & 40\% & 44Maler & 803586351807479809 & 40\%\\
KumlehnLisa & 803584267653709824 & 39\% & FlorianWanken & 803586889802465280 & 47\%\\
DreysKatharina & 803580625911480320 & 49\% & IHeulach & 803917039597473792 & 47\%\\
LaurySamsy & 803575433862283264	& 53\% & porryflo12 & 803915544755847168	& 48\%\\
Eva\_Omaha & 803577951040180224 & 46\% & DiamondGirl\_97 & 803595282969755648 & 40\%\\
Jonas\_Der\_Baum & 803581393393643520 & 37\% & kenny\_boy300 & 803278016709332994 & 42\%\\
NickyTheMan1 & 803580666260705281 & 42\% & Saschamachtsgut & 803279247804628992 & 41\%\\
ruediwig & 803579187344904192 & 56\% & ollerbaum121 & 803574359281598464 & 52\%\\
Kalle\_dod & 803271431400394752 & 54\% & DinoDingi & 803582934376738816 & 60\%\\
Shagggy\_93 & 803274850768920576 & 46\% & wernerbbbright & 803584898271444992 & 48\%\\
Luise\_D2 & 803583207069351936 & 58\% & The\_pfist & 803881433840361472 & 47\%\\
hoppendorf & 803584623594897409 & 51\% & hansemeister11 & 803901587827621889 & 54\%\\
sabinepeterson7 & 803585050721783808 & 57\% & wendtneraxxxy & 803912137630420992 & 56\%\\
ullaschoene80 & 803988474416295937 & 54\% \\\hline
\end{tabular}
\end{table}


\begin{figure}[htb]
	\centering
    \includegraphics[width=.4\textwidth]{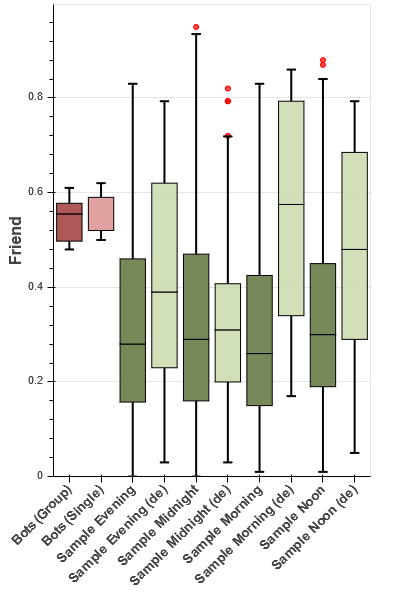}
    \includegraphics[width=.4\textwidth]{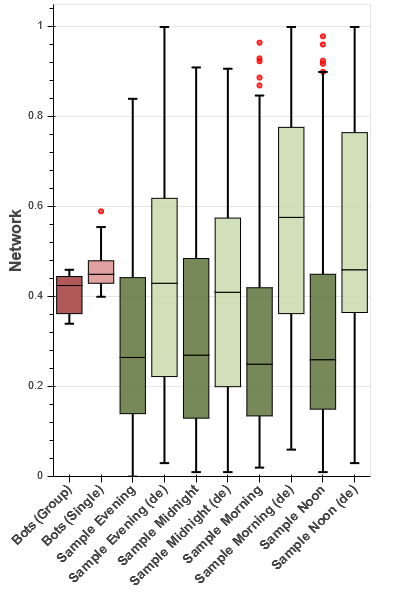}
    \includegraphics[width=.4\textwidth]{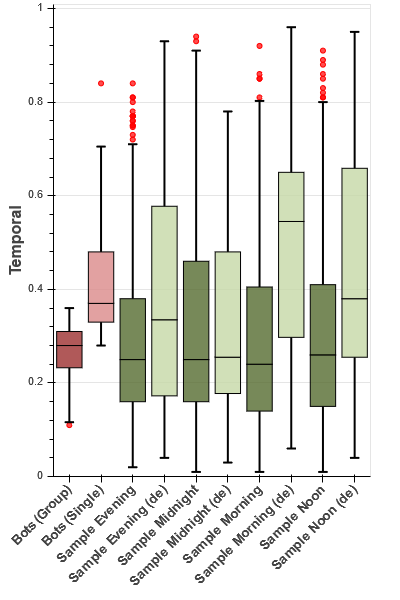}
    \includegraphics[width=.4\textwidth]{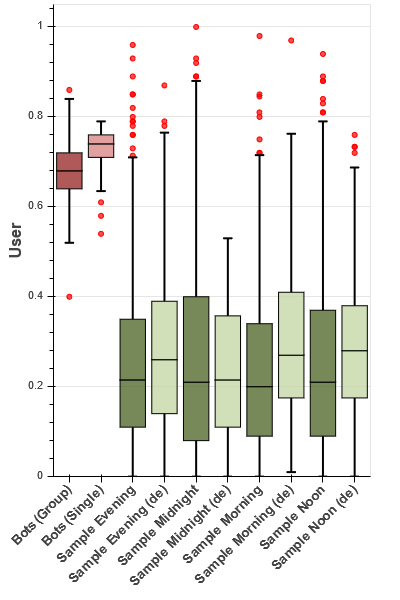}
    \caption{\label{fig:bot_score_detail2}Detailed statistics of four meta features (friendship, network, temporal, and user) for our social bots (red) and the baseline accounts worldwide (green) and from Germany (light green).}
\end{figure}

\appendix

\bibliography{socialbots}

\section{Code of a Simple Twitter Bot}
\label{sec:simple_bot_listing}
The following code is a fully functional Twitter bot which continuously tracks the Twitter stream for a given hash tag (\texttt{\#Hashtag}) and instantly replies to the sender with a simple 'Hello'. Please note, that login information for the Twitter API has been obscured and some error handling code has been removed for brevity. 

{\scriptsize
\pythonexternal{SimpleBot2.py}
}

\end{document}